\documentclass[useAMS,usenatbib]{mn2e}
\usepackage[pdftex]{graphicx}
\usepackage{color}

\def\be{\begin{equation}}
\def\ee{\end{equation}}
\def\bea{\begin{eqnarray}}
\def\eea{\end{eqnarray}}

\newcommand{\htwo}{${\rm H_2}$}
\newcommand{\hsix}{${\rm H_6^+}$}
\newcommand{\hd}{${\rm (HD)_3^+}$}
\newcommand{\ph}{{\it p}-H$_2$}
\newcommand{\percm}{${\rm cm^{-1}}$}
\newcommand{\mum}{$\,{\rm\mu m}$}

\newcommand{\alert}[1]{\textcolor{black}{#1}}

\title[Dielectric function of solid H$_2$]{The infrared dielectric function of solid \alert{para-}hydrogen}
\author[Kettwich et al]{Sharon$\!$ C.$\!$ Kettwich,$^{\!\!1,2}$
David$\!$ T.$\!$ Anderson,$^{\!\!2}$\thanks{danderso@uwyo.edu}
Mark$\!$ A.$\!$ Walker$^{3}$\thanks{Mark.Walker@manlyastrophysics.org}
Artem$\!$ V.$\!$ Tuntsov,$^{\!\!3}$\\ 
$\!$1. Department of Chemistry and Physical Science, Temple College, 2600 S. 1st St., Temple, TX 76504 , U.S.A.\\
$\!\!$2. Department of Chemistry, University of Wyoming, Laramie, WY 82071, U.S.A.\\
$\!\!$3. Manly Astrophysics, 3/22 Cliff St,  Manly 2095, Australia}

\begin{document}

\date{Accepted: {\it 18th March 2015.\/} Received: {\it 16th February 2015.\/} In original form: {\it 20th October 2014.\/}}

\pagerange{\pageref{firstpage}--\pageref{lastpage}} \pubyear{2014}

\maketitle

\label{firstpage}

\begin{abstract}
We report laboratory measurements of the absorption coefficient of solid {\it para\/}-\htwo, within the wavelength range from 1 to 16.7$\,{\rm\mu m}$, at high spectral resolution. In addition to the narrow rovibrational lines of \htwo\ which are familiar from gas phase spectroscopy, the data manifest double transitions and broad phonon branches that are characteristic specifically of hydrogen in the solid phase. These transitions are of interest because they provide a spectral signature which is independent of the impurity content of the matrix. We have used our data, in combination with a model of the ultraviolet absorptions of the H$_2$ molecule, to construct the dielectric function of solid {\it para\/}-H$_2$ over a broad range of frequencies. Our results will be useful in determining the electromagnetic response of small particles of solid hydrogen. The dielectric function makes it clear that pure \htwo\ dust would contribute to IR extinction predominantly by scattering starlight, rather than absorbing it, and the characteristic IR absorption spectrum of the hydrogen matrix itself will be difficult to observe. 
\end{abstract}

\begin{keywords}
ISM: molecules --- dust, extinction --- molecular processes --- scattering
\end{keywords}

\section{Introduction}
As a candidate material for interstellar dust, solid hydrogen is not new. But it is not well explored. Shortly after being proposed as a possible dust constituent (e.g. Wickramasinghe and Reddish 1968), it was recognised that the pure solid would sublimate rapidly under typical interstellar conditions (Field 1969; Greenberg and de Jong 1969), and \alert{interest in the material soon waned.  Subsequent studies of ``condensed'' \htwo\ have been predominantly concerned with the distinct case of \htwo\ molecules adsorbed onto other materials --- either individual atoms/molecules/ions (e.g. Duley 1996; Bernstein, Clark and Lynch 2013), or else the surfaces or porous matrices of other solids (e.g. Sandford and Allamandola 1993). Such composites differ from bulk solid \htwo\ in many ways, including the spectroscopic properties of the hydrogen -- e.g the Q$_1$ line, which was the main focus of Sandford and Allamandola (1993), is entirely absent in the bulk (see \S2 of this paper) -- and consequently we do not consider them here.}

\alert{Pfenniger, Combes and Martinet (1994) suggested that spiral galaxies might contain gas in cold, dense, molecular clouds which have escaped direct detection --- i.e. a form of baryonic dark matter. In such clouds the \htwo\ could reach its saturated vapour pressure, allowing pure solid hydrogen to precipitate (Pfenniger and Combes 1994; Wardle and Walker 1999; Pfenniger 2004).  Some aspects of those models are controversial, but not the following point: any gas which is cold and dense enough to allow \htwo\ to precipitate would likely have escaped direct detection to date. The study of solid H$_2$ formation sites is, therefore, primarily a theoretical endeavour at present. But from those formation sites hydrogen grains can be injected into the broader ISM (e.g. by continuous stripping of surface material, or by disruptive events), where they may be more readily observed.} Furthermore, we now know that the thermodynamics of the pure material are misleading: in interstellar space, solid H$_2$ would acquire a dense layer of surface charges, and these make hydrogen grains much more durable (Walker 2013). Given that it may be possible for hydrogen dust to survive, sheer weight of numbers, relative to any of the metals, motivates an exploration of what such particles might look like.

The molecular ion \hsix\ is one aspect of that appearance. It does not feature in the ionisation chemistry of H$_2$ in the gas-phase, but is known to form in the irradiated solid (Symons \& Woolley 2000; Suter et al 2001; Kumada, Tachikawa and Takayanagi 2005; Kumada, Takayanagi and Kumagai 2006; Kumagai et al 2007); as such it provides a signature of the condensed phase. The vibrational lines of this species, and its isotopic variant \hd, were characterised by Lin, Gilbert and Walker (2011), who reported a striking match to the pattern of strong infrared bands of the interstellar medium (ISM). 

A thorough investigation of the properties of \htwo\ dust is called for, and for that we need to establish the dielectric function of pure, solid \htwo. This paper presents a preliminary description of that dielectric function. Our model is based on our own infrared absorption measurements of the solid in the wavenumber\footnote{\alert{Throughout this paper we employ wavenumbers measured in \percm, i.e. just the reciprocal of the wavelength measured in cm.}} range 600-10{,}000$\;$\percm\ \alert{(16.7-1\mum),} supplemented by published data on the lowest rotational transition of the solid, which lies outside the range of our spectrometer.  In the case of isolated \htwo\ molecules the infrared lines are electric quadrupole transitions, and consequently the absorptions we have measured are weak compared to those in the UV, which correspond to electric dipole transitions of the individual molecules. Although we have obtained no UV spectra, the strength of those transitions makes it essential to include them in any model of the dielectric properties. We have therefore represented the discrete UV transitions with Lorentz oscillators and the bound-free continuum with a published analytic fit to the measured absorption cross-section of the \htwo\ molecule. We then synthesised the dielectric function via the Kramers-Kronig relations.

Although our model of the UV absorptions is incomplete, the omissions are not very significant in respect of the overall dielectric response --- the model captures 95\% of the oscillator strength, for example. However, it is not a good description of the detailed structure of the UV absorption bands of the solid, because it is based on the measured response of gas-phase molecules and then simply scaled to the density of the solid. This approach fails to account for the line broadening which arises from intermolecular interactions within the solid. Away from the UV absorption bands themselves, our model should be reliable. In particular it should give a satisfactory account of the influence which the strong UV bands have on the infrared dielectric function, and it thus provides appropriate context for our infrared measurements of the solid itself. Despite the shortcomings of our dielectric function in the UV, we present the complete model here. The motivation for doing so is simply that a better model is not available at present. If it is used in the UV, it should be used with caution.

The structure of this paper is as follows. In the next section we describe our laboratory techniques and the infrared (IR) absorptions of solid {\it para\/}-\htwo\ which we have measured.\footnote{The {\it ortho/para\/} designation refers to odd/even angular momentum quantum numbers. Because {\it ortho-para\/} state transitions are forbidden, conversion between the two forms is slow, and these sequences constitute separate populations in statistical equilibrium. However, at low temperatures all the molecules eventually end up in the $J=0$ state, so it is the {\it para\/} form which is relevant here. Henceforth we use the shorthand \ph\ for {\it para\/}-\htwo.} In section 3 we present our model of the UV response of the solid, based on the discrete and bound-free transitions of \htwo\ reported in the literature, and this is combined with our IR absorption data to yield a single description of the dielectric function. Discussion and conclusions follow in \S\S4,5.

\section{Laboratory measurements of infrared absorption}
At low temperatures, infrared absorption spectra of \ph\ in the gas phase are simple, consisting of the pure rotational line ${\rm S_0(0)\/}$, at approximately $354\,$\percm\ \alert{(28.2\mum),} and a sequence of rovibrational lines ${\rm S}_v(0)$, with $v=1,2,3\dots$. Here, and subsequently, we follow the usual spectroscopic nomenclature in which changes in the angular momentum quantum number by amounts $\Delta J=-1,0,1,2,4$ are  denoted by the letters ${\rm P, Q, R, S, U}$, respectively. The subscript on the designation indicates the vibrational quantum number of the upper state, and the number in parentheses denotes the rotational quantum number of the lower state, which is always $0$ for \ph\ at low temperatures. Absorptions to vibrational states higher than $v=1$ arise from anharmonicity in the interatomic potential and are comparatively weak, with oscillator strengths that decrease by a factor $\sim10$ for each of the first few steps up the vibrational ladder  (Turner, Kirby-Docken and Dalgarno 1977). 

Once \ph\ molecules are condensed into the solid phase, however, much more structure is seen (e.g. Gush et al 1960; Buontempo et al 1982; Mengel, Winnewisser and Winnewisser 1998; Mishra et al 2004). For example, a ${\rm U_1(0)}$ line appears with significant intensity, induced by intermolecular interactions of a central \ph\ molecule with the surrounding \ph\ matrix\footnote{The neighbouring molecules each have a non-zero electric quadrupole moment.} (Steinhoff et al 1994). In addition to transitions associated with a single molecule, one sees double transitions, e.g. ${\rm Q_1(0)+S_0(0)}$, in which two neighbouring molecules make a simultaneous change in quantum state. Triple transitions have also been identified (Mengel, Winnewisser and Winnewisser 1998) but they are much less prominent. And each transition manifests a broad sideband, on the short wavelength side, in which phonons carry off energy into the lattice. Phonon-assisted transitions are designated with the subscript ${\rm R}$. Some transitions which are strictly forbidden in isolated \ph\ molecules can, with phonon assistance, acquire substantial intensity in the solid. A good example is the ${\rm Q_R(0)}$ band, which is prominent in solid \ph, despite the fact that $J=0\rightarrow J=0$ is a forbidden transition (Field, Somerville and Dressler 1966); and the ${\rm Q}_1(0)$ line itself is absent even in the solid (Hinde 2003; Anderson et al 2002).

A theoretical basis for describing the infrared specrum of solid \ph\ was pioneered by Van~Kranendonk and collaborators (e.g. Van Kranendonk and Karl 1968). Many of the observed features can be understood within a framework where the quanta of molecular rotation and vibration are to some degree delocalised in the lattice, and the terms ``roton" and ``vibron" are applied to those lattice excitations.

Those lines which correspond to transitions of individual molecules are seen to occur at slightly different wavelengths in the solid --- a ``matrix shift''. For example in solid \ph\  the ${\rm S_1(0)\/}$  line appears at approximately $4{,}486\,$\percm\ \alert{(2.229\mum),} roughly $12\,$\percm\ lower in wavenumber than the same line in the gas phase. The IR transitions of solid \ph\ are typically much more intense than the analogous gas phase transitions, which occur by quadrupole radiation, because the transitions acquire some electric dipole character via intermolecular interactions in pairs of \htwo\ molecules --- analogous to collision-induced IR activity (Raston and Anderson 2007).

\subsection{Experimental method}
We prepared \ph\  crystals using the rapid vapor deposition (RVD) method of Fajardo and Tam (1998).  In this method the crystal is grown by deposition of pre-cooled \ph\  gas onto a \alert{2$\,$mm thick} BaF$_2$ optical substrate held at $\sim2.5\;$K within a sample-in-vacuum liquid-He cryostat.  \alert{In order to minimise the effects of high-order internal reflections in the optical substrate, its front and back surfaces are not parallel; the wedge angle is 1$^\circ$. For the same reason, we probe the sample with a weakly converging IR beam rather than a collimated beam.} The \ph\  gas is enriched from {\it normal\/}-\htwo\ gas to greater than 99.97\% \ph\  using a variable temperature {\it ortho/para\/} converter operated near $14.0\;$K.  The {\it ortho\/}-\htwo\ concentration in the sample can be determined using the ${\rm Q_1(1)}$ or ${\rm U_1(1)}$ transitions and their known absorption strengths.

The vacuum chamber that surrounds the cold optical substrate has a nominal pressure of $10^{-6}\;$Torr prior to any cryogens being introduced.  This is sufficiently large that all samples become increasingly contaminated with \htwo O and CO$_2$ impurities over the lifetime of the crystal (approximately 4 to 10 hours).  These trace impurities do not cause any great difficulty for us, however, as most of their absorption bands do not overlap with those of the \ph\ itself.

High-resolution Fourier Transform IR (FTIR) spectroscopy was performed on each sample using a normal-incidence transmission optical setup.  The IR beam is focused through the sample with off-axis parabolic mirrors of 8$^{\prime\prime}$ effective focal length.  The optical path outside the cryostat and spectrometer is contained within an optical conduit purged with dry air to minimize atmospheric absorptions.  FTIR spectra were recorded at various resolutions ranging from 0.01 to 0.1 \percm\ both during and after deposition.  The FTIR spectrometer (Bruker IFS 120HR) used to record IR spectra is equipped with a variety of sources, beamsplitters and detectors to cover the full range of the spectrum.  Mid-IR spectra are recorded with a glowbar source, a Ge-coated KBr beamsplitter, and a liquid nitrogen cooled HgCdTe detector in the region from 600 to 6{,}000 \percm\ at 0.1~\percm\ resolution, and with an InSb detector from 2{,}000 to 5{,}000~\percm\ at 0.09~\percm\ resolution.  Near-IR spectra are recorded with a tungsten source, a CaF$_2$ beamsplitter, and a liquid nitrogen cooled InSb detector in the region from 2{,}000 to 10{,}000~\percm\ at 0.09~\percm\ resolution.

The three instrumental configurations just described provide high resolution spectra over a broad wavelength range, including the fundamental and first overtone vibrational modes. Unfortunately, our spectral coverage does not include the lowest pure rotational transition, i.e. ${\rm S_0(0)}$, nor the peak of the associated phonon branch ${\rm S_R(0)}$

\subsection{Thickness determination}
In their original paper in 2001, Tam and Fajardo measured the integrated absorption intensities of the ${\rm Q_1(0)+S_0(0)}$ ($\sim$4{,}508~\percm) and ${\rm S_1(0)+S_0(0)}$ ($\sim$4{,}840~\percm) double transitions of \ph\ and calibrated them against the sample thickness determined interferometrically.  They also showed that the integrated intensities of these double transitions are insensitive to crystal structure and the presence of impurities. Furthermore the ${\rm Q_1(0)+S_0(0)}$ transition is approximately ten times stronger than the ${\rm S_1(0)+S_0(0)}$ transition, so one can use the stronger transition for thin samples ($d\la0.02\,$cm), the weaker one for thick samples (up to 10~cm), and the combination for intermediate thicknesses, allowing for consistency checks.  We have therefore followed Tam and Fajardo (2001) and used these double transitions to determine the thickness of our samples.

\begin{figure}
\includegraphics[width=85mm]{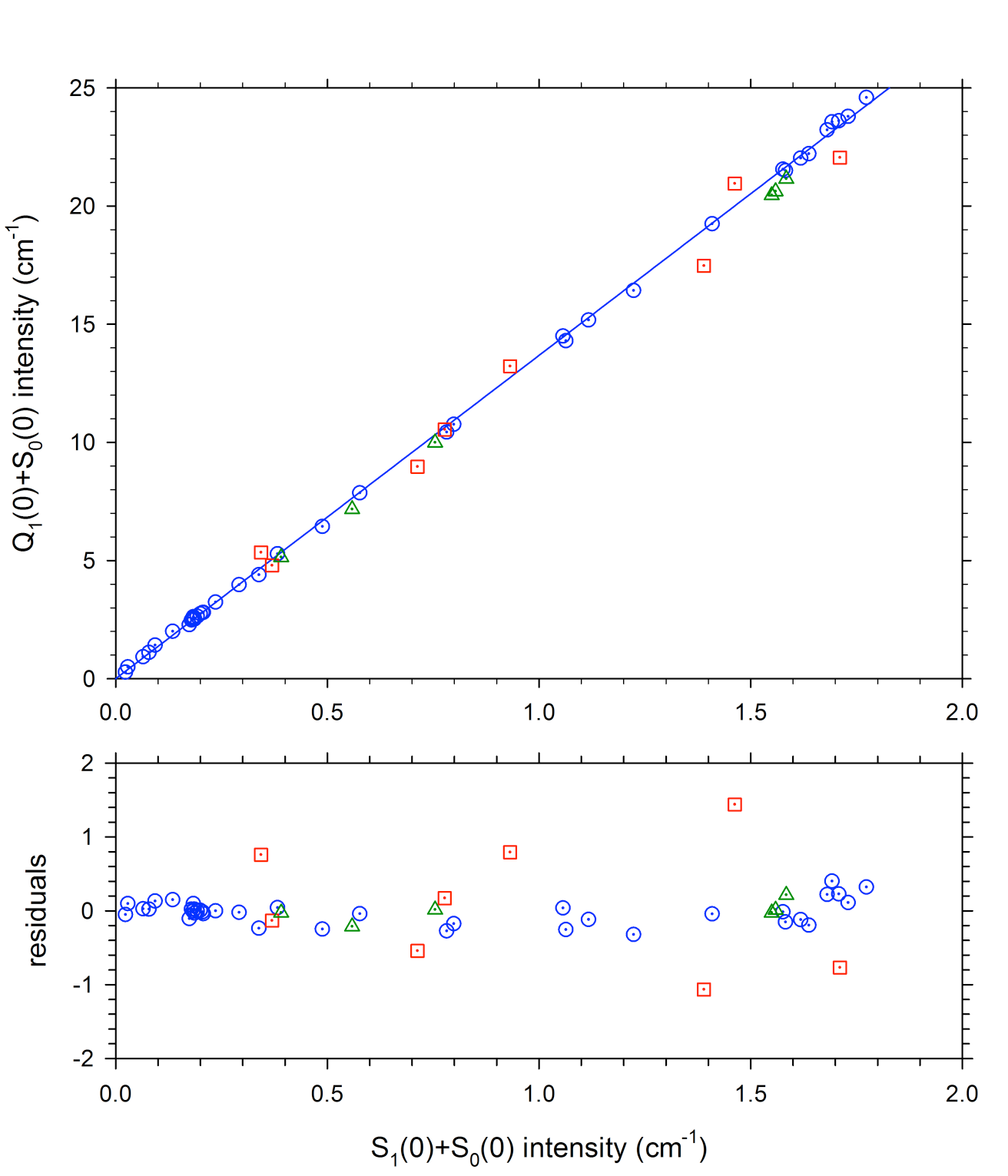}
\vskip-0.1truecm
\caption{Upper panel: integrated intensity of the ${\rm Q_1(0)+S_0(0)}$ transition, versus that of the ${\rm S_1(0)+S_0(0)}$ transition of solid \ph, as measured with the three beam-splitter/detector combinations used in the present study. Blue circles denote measurements made with CaF$_2$/InSb; green triangles for KBr/InSb; and red squares for KBr/HgCdTe. The plotted line is the best-fit relationship, in a least-squares sense, of a proportionality between the two integrated intensities. Lower panel: residuals relative to the best-ft relationship.}
\end{figure}

To check the internal consistency of this method,  we examined the correlation between the integrated intensities of these double transitions determined using all three of our beam-splitter/detector combinations; results are shown in Figure 1.  Both FTIR setups that use the KBr beam-splitter showed a slope consistent with the previous measurements of Tam and Fajardo (2001).  However, for the setup that employed a CaF$_2$ beam-splitter, the determined slope was approximately 6\% larger than the literature value and outside the 95\% confidence range.  The reasons for this discrepancy are unclear. However, to determine the thickness using a common protocol over the entire spectral range we decided to use the integrated intensity of the ${\rm Q_1(0)+S_0(0)}$ double transition of $90 (\pm2)\;{\rm cm^{-2}}$ (Fajardo 2011), and to scale the integrated intensity of the ${\rm S_1(0)+S_0(0)}$ accordingly for the three setups.

\begin{figure}
\includegraphics[width=85mm]{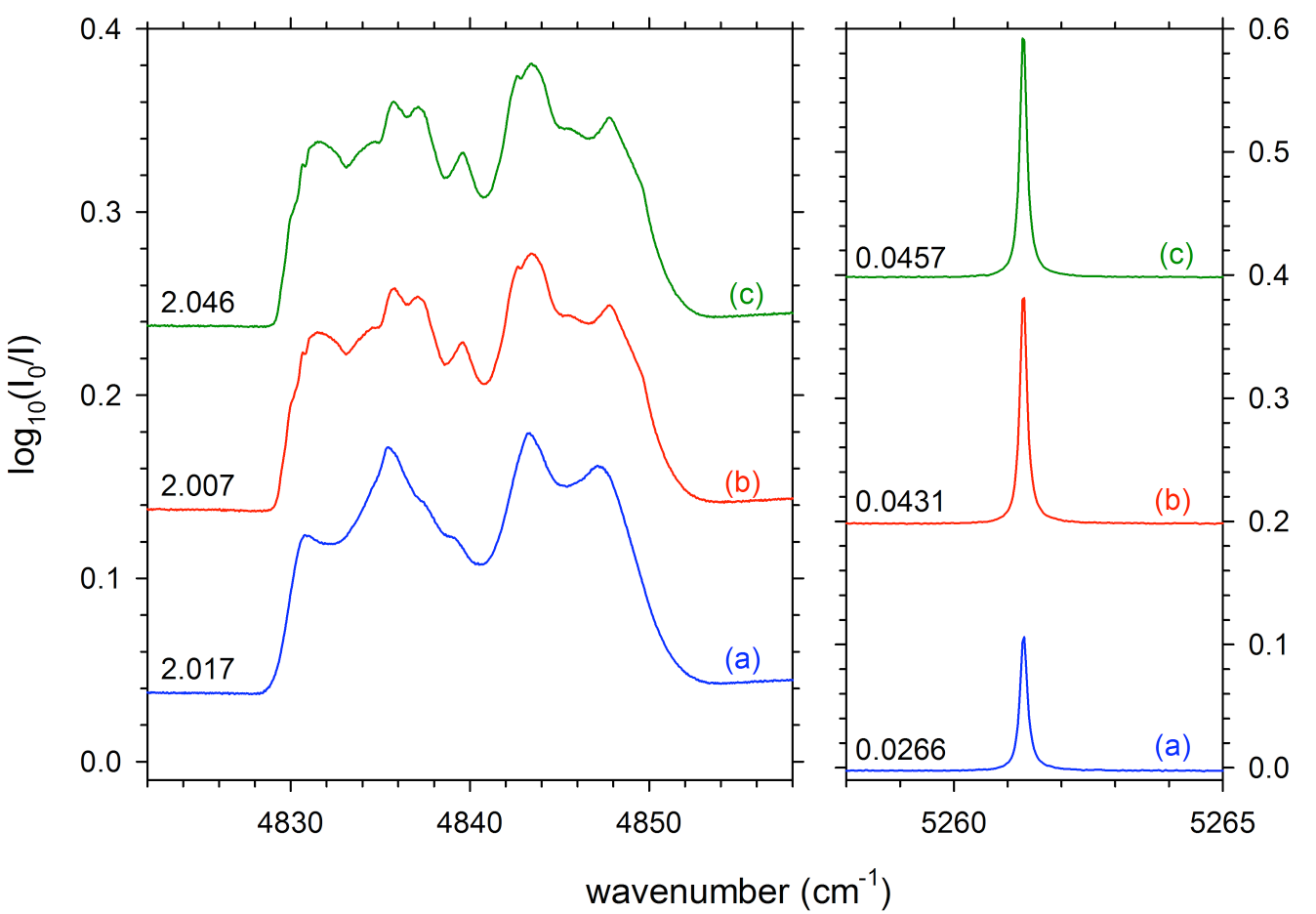}
\vskip-0.1truecm
\caption{Line profiles for the ${\rm S_1(0)+S_0(0)}$ double transition (left panel), and the ${\rm U_1(0)}$ single transition (right panel), of solid \ph. These profiles are taken from spectra of the same sample (approx 3~mm thick), recorded at different times during the experiment. Trace (a) was recorded at 1.7~K, in the freshly deposited sample; trace (b) during annealing of the sample, for 30 minutes at 4.3~K; and trace (c) after annealing and re-cooling the sample to 1.6~K. The numerical value adjacent to each trace is the integrated intensity of each line (in \percm). The double transition exhibits some profile evolution, but its integrated intensity is steady to within $\sim1$\%. By contrast the integrated intensity of the single transition increases substantially during annealing, as fcc crystal domains are converted to the more thermodynamically stable hcp structure.}
\end{figure}

\subsection{Annealing of samples}
Samples of \ph\  prepared by RVD contain a mixture of face centered cubic (fcc) and hexagonal close packed (hcp) structures in the freshly deposited material.  The fcc crystal domains are, however, only metastable: they arise because the rapid deposition of material does not allow time for relaxation into the thermodynamically more stable hcp structure. Although a freshly deposited solid \ph\ sample can contain upwards of 50\% in fcc domains, annealing of the sample at 4.3~K for a period of several minutes will almost completely convert the fcc into hcp structures. \alert{In an astrophysical context there is plenty of time for solid \ph\  to relax into its minimum energy configuration, so it is the more stable, hcp structure that is relevant, and consequently all of our samples underwent annealing.} Spectra were recorded during deposition and annealing, and after the annealing cycle was complete.

Although the fcc and hcp structures of solid \ph\  are very similar, their spectra are not identical. A key difference is that in the fcc structure, but not in the hcp structure, each \ph\ molecule occupies a crystal site with center-of-inversion symmetry, so that the induced dipole moment for the \ph\ molecule in this site cancels out to a large degree for the fcc structure. Consequently some of the single transitions of solid \ph\ are much stronger in hcp than in fcc; a good example of this is the ${\rm U_1(0)}$ transition, which is only induced in hcp crystal sites (Raston et al 2013).

Considerations of the symmetry of an individual site in the lattice are less relevant to the strength of double transitions, and those transitions may exhibit integrated intensities which are independent of the fcc/hcp content of the sample. That is one reason why double transistions are preferred for thickness determination --- as described in \S2.2. Figure 2  illustrates both the ${\rm S_1(0)+S_0(0)}$ and the ${\rm U_1(0)}$ transitions, recorded simultaneously in a single sample: right after deposition was complete (trace a), while annealing the sample at 4.3~K for 30 minutes (trace b), and after re-cooling the sample to 1.57~K (trace c). In detail the profile of the ${\rm S_1(0)+S_0(0)}$  double transition is seen to change slightly between the as-deposited, and annealing or annealed spectra; but the total intensity, integrated over the profile, is constant to within $\pm1$\%. By contrast the ${\rm U_1(0)}$ transition increases in strength during annealing, and is 70\% stronger in the annealed sample compared to the freshly deposited material.

\subsection{Measured absorption coefficients}
From the measured intensity, $I(\nu)$, we can determine  the absorption coefficient, $\alpha(\nu)$, which is a material property and thus independent of the sample thickness, $d$. The two are related by
\begin{equation}
\alpha(\nu)\;\alert{\simeq}\;{1\over d}\log_e\!\left[{{I_0(\nu)}\over{I(\nu)}}\right],
\end{equation}
where $I_0(\nu)$ is the measured intensity in the absence of a \ph\ sample. \alert{Equation 1 would be exact in the case of a material with unit refractive index. As we will see in due course, solid \ph\ has a refractive index that is very close to one throughout the range of our measurements, so the approximation in equation 1 is a good one.} For the purpose of deriving absorption coefficients, we have used only spectra from fully annealed crystals of \ph.

Figure 3 shows our measured absorption coefficients in three distinct spectral regions. The regions around the fundamental and first vibrational overtone exhibit the huge range of linewidths that is one of the remarkable features of the IR spectra of solid hydrogen. For example, the sharp transition ${\rm U_1(0)}$ at 5{,}261$\,$\percm \alert{(1.901\mum)} has a FWHM $\simeq0.16\,$\percm, whereas the phonon-assisted transition ${\rm Q}_R(0)$, which has a similar peak absorption coefficient (located near 4{,}230$\,$\percm), displays a FWHM in excess of $100\,$\percm.

The ${\rm U_1(0)}$ excitation is essentially localised on a single molecule, because the vibrational energy of this transition lies below the \ph\ vibron band. In turn that is because of centrifugal distortion of the \htwo\ molecule in the $J=4$ state.

\begin{figure}
\hskip-0.1cm\includegraphics[width=85mm]{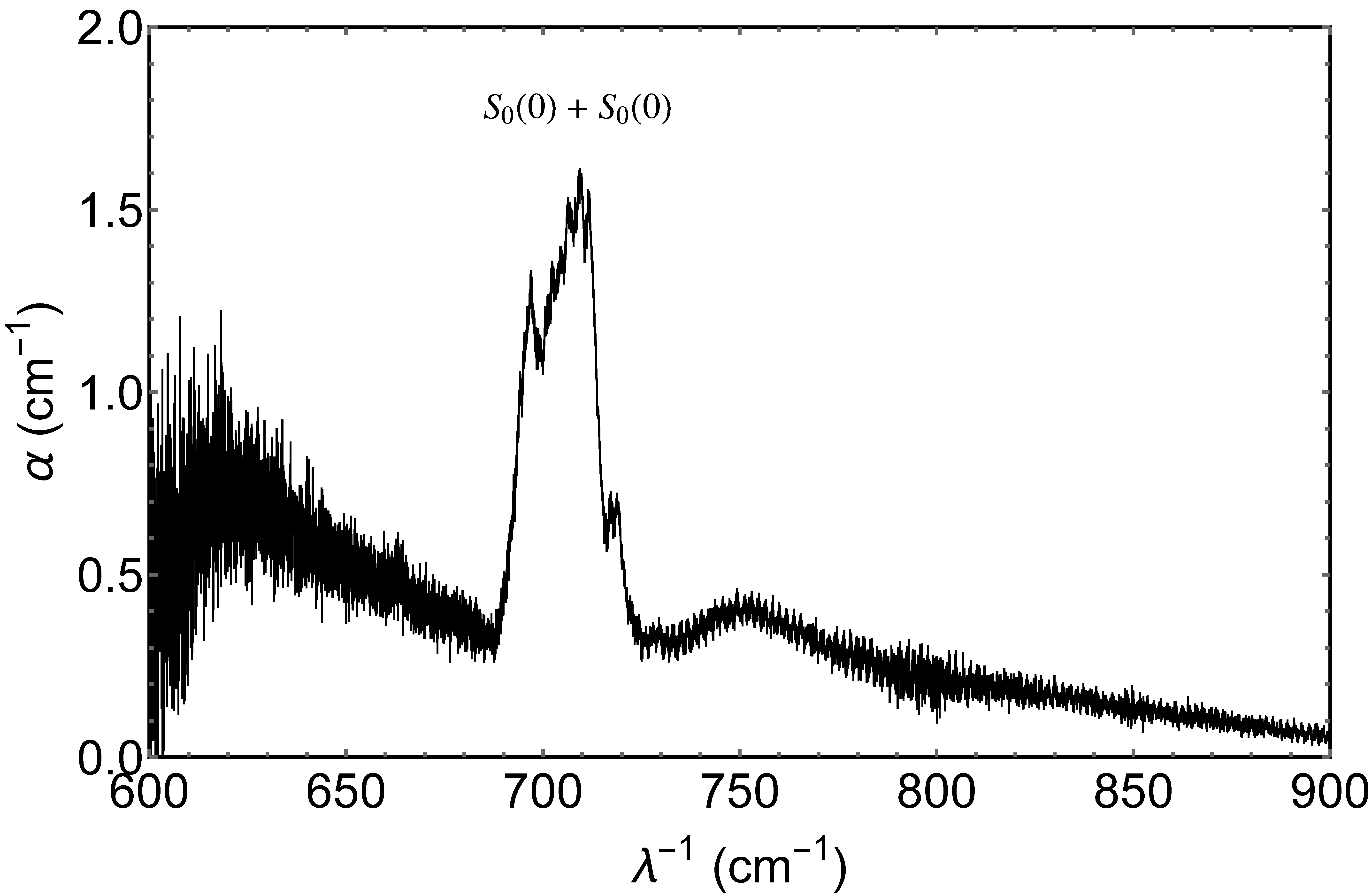}
\vskip-0.1truecm
\hskip0.3cm\includegraphics[width=82mm]{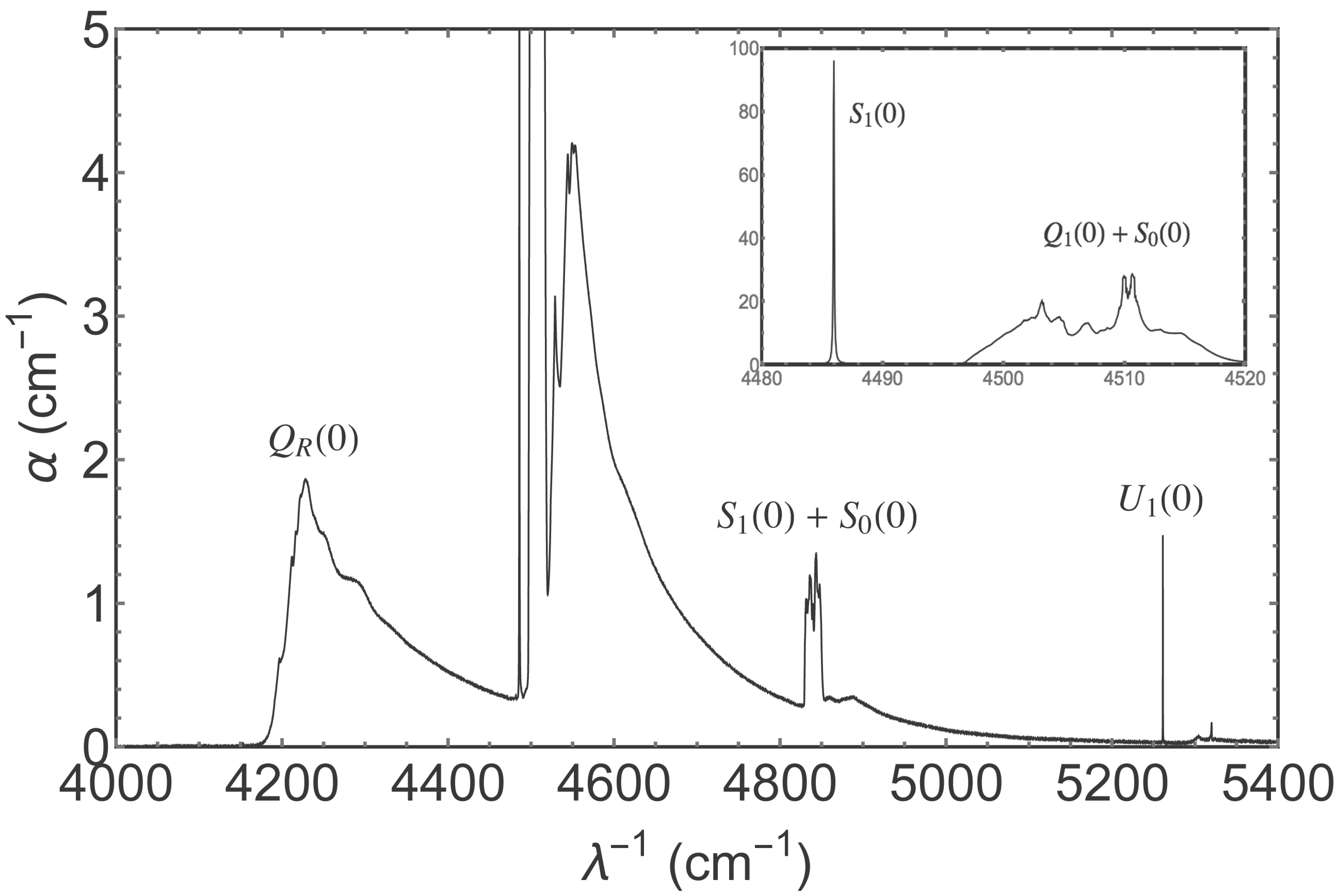}
\vskip-0.1truecm
\includegraphics[width=85mm]{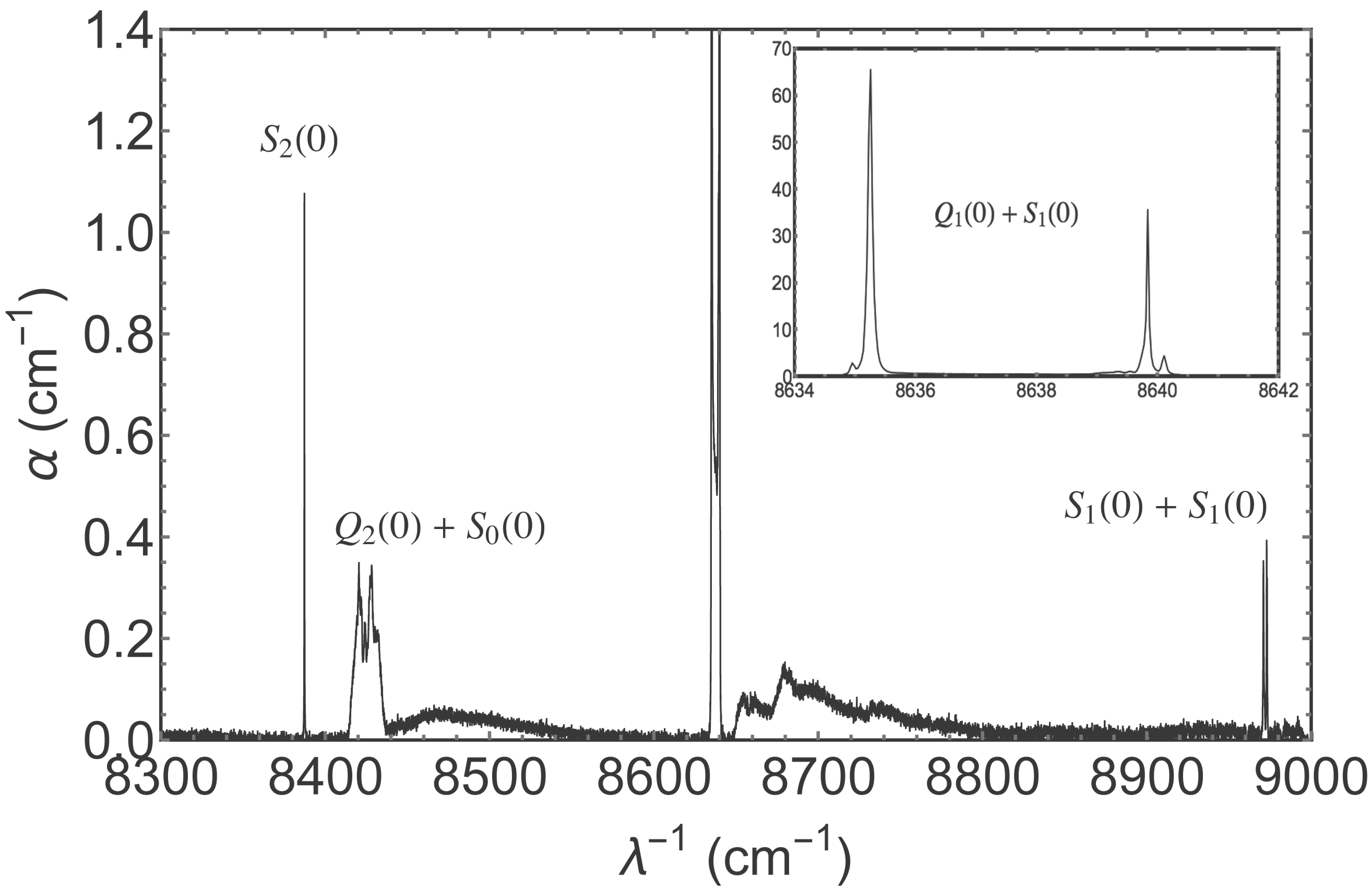}
\vskip-0.1truecm
\caption{The measured absorption coefficient of solid \ph, showing: pure rotational transitions (top panel); the fundamental vibrational band (middle panel, \alert{with inset showing the full range of the ${\rm S_1(0)}$ and ${\rm Q_1(0)+S_0(0)}$ absorptions}); and the first vibrational overtone (bottom panel, \alert{with inset showing the full range of the ${\rm Q_1(0)+S_1(0)}$ absorption}). Assignments are shown for the most prominent transitions. All zero-phonon absorption lines are accompanied by a broad phonon branch at slightly higher wavenumber. Note also the strong ${\rm Q_R(0)}$ phonon branch, peaking at approximately 4{,}230$\,$\percm, but no evidence of the forbidden ${\rm Q_1(0)}$ line itself (near 4{,}150$\,$\percm). The continuum absorption seen below 700$\,$\percm\ is the phonon branch of the ${\rm S_0(0)}$ line, which lies beyond the $\lambda^{-1}\sim600\,$\percm\ instrumental limit. No wavelet filtering has been applied to the data at this stage.}
\end{figure}

The ${\rm U_1(0)}$ transition has its own phonon side-band which is broad and peaks roughly $45\,$\percm\  higher in wavenumber than the zero-phonon line. This phonon band overlaps with the double transition ${\rm Q_1(0)+U_0(0)}$, which is the source of the sharp peak at 5{,}320$\,$\percm\ \alert{(1.880\mum).} Broad, asymmetric, phonon side-bands can be seen to be situated in a similar relationship to essentially all of the ro-vibrational transitions -- both single and double -- in figure 3. 

The double transitions have complicated line profiles and may themselves be quite broad (i.e. even the zero-phonon transition), as can be seen with the ${\rm S_1(0)+S_0(0)}$ transition in figures 2 and 3.

It is no surprise that the transitions in the vibrational overtone region are generally weaker than those of the fundamental band, but some significant peaks in the absorption coefficient are nevertheless manifest in the bottom panel of figure 3. Anharmonicity of the \htwo\ molecule is evident in the displacement of the ${\rm Q_2(0)+S_0(0)}$ transition, relative to the ${\rm Q_1(0)+S_1(0)}$ transition, of more than 200$\,$\percm.

\subsection{Data processing}
Although our laboratory spectra provide a wealth of information on the absorption coefficient of solid \ph, the data in their raw form are not ideal for constructing a model dielectric function. In this section we describe how we used the data to arrive at a model for the infrared properties of the solid.

\subsubsection{Saturated data}
Our principal dataset, obtained for the express purpose of constructing the dielectric function of solid \ph, consists of two spectral scans, covering the range 2{,}000-10{,}000$\,$\percm, of a solid \ph\ crystal of thickness 3.1~mm. In these data we observed that two \ph\ features -- the  ${\rm S_1(0)}$ line and the sharper peak of the ${\rm Q_1(0)+S_1(0)}$ transistion -- are so strong and narrow that they are saturated. We anticipated this difficulty and  therefore recorded similar scans of a crystal of thickness 0.31~mm, in which these lines are not saturated. Using the spectra of the thinner crystal it was straightforward to determine the line profiles in the narrow regions where saturation is a problem in our main dataset. Both regions are well represented by Lorentzian profiles with FWHM of approximately $0.08\,$\percm. The corresponding model profiles were used to replace the saturated points in our main dataset.

\subsubsection{Contaminants}
As noted earlier, our samples of solid \ph\ exhibit a low level of contamination, notably from  \htwo O and CO$_2$, and these impurities contribute to the absorption spectra. However, the spectral features of these molecules are well known and thus easily distinguished from those of \ph. Mostly their contributions lie in regions where the \ph\ absorption is in any case too small to be measured in our experiments, and in those regions we simply do not use the data so the contamination is irrelevant. The exception is a combination line of water at 5{,}338.54$\,$\percm, which is superimposed on the phonon continuum associated with the ${\rm U_1(0)}$ and ${\rm Q_1(0)+U_0(0)}$  transitions. This water line is well reproduced by a Lorentz oscillator, and we therefore removed the feature by subtracting the best-fit Lorentzian profile. 

In addition to the low-level contamination of our crystals, just described, the spectrometer itself -- specifically, the HgCdTe detector -- has, over the years, acquired a surface contaminant. The nature of this contamination is unknown, but the effect on our acquired spectra is to create strong non-linearities, and thus large artefacts in the low-frequency absorption coefficients. In practice this meant that the low-frequency spectra we acquired specifically to measure the absorption coefficient were not useful. Instead we made use of some archived spectra from a study of acetonitrile (${\rm CH_3CN}$), at a concentration of 23~ppm, in a solid \ph\ matrix. These spectra were taken before significant detector contamination had built-up. \alert{Strong, narrow ($\sim0.1\,$\percm) acetonitrile absorptions are present in these spectra --- e.g. at 919$\,$\percm\ and 1041$\,$\percm. But from the known gas-phase spectra of this molecule (Nishio et al 1995), the only contaminating line expected within the portion of data which we make use of is an overtone line at approximately 718$\,$\percm. There is no sign of a narrow absorption feature at this location in our data, so we conclude that it is at or below the level of the noise in the data, and therefore insignificant.}

\subsubsection{Wavelet filtering}
\alert{Our absorption coefficient spectra, as recorded, include some low-amplitude, high frequency fringes --- evident around 750$\,$\percm\ in figure 3. These fringes are very likely a residual systematic calibration error arising from multiple internal reflections in the BaF$_2$ substrate.} Our spectra, particularly the top and bottom panels of figure 3, also exhibit significant measurement noise. Where there is strong absorption the measurement noise is unimportant, but in spectral regions where the absorption is weak the signal we're interested in may be masked by the noise. Furthermore, the presence of noise causes numerical difficulties in constructing the real part of the refractive index, via the Kramers-Kronig relations, because one expends a great deal of time carefully integrating all the details of the noise in the vicinity of the point of interest.

To address these problems we ``denoised'' our absorption coefficient spectra using wavelet transforms. Wavelet filtering is preferred over spectral averaging because the wavelet approach preserves the profile of strong, narrow features in the data, whereas spectral averaging does not.

Within the {\it Mathematica\/} software package, we utilised the discrete wavelet transform, with a Haar wavelet; we zeroed all coefficients below a suitably chosen threshold, and then applied the inverse discrete wavelet transform. The thresholding value was chosen to be close to the level of the noise in the data, so as to filter out the noise while preserving the signal in so far as possible. In regions where the absorption is small, the filtered spectra are piecewise-constant approximations to the true, smoothly varying coefficient.

\subsubsection{Model phonon band}
In regions where the absorption coefficient of solid \ph\ is very small indeed, even our ``denoised'' data do not yield an accurate value. In this case there is little merit in using the data themselves, and instead we have employed a simple model of the continuum absorption based on the observed properties of the phonon side-bands. 

In our data  ${\rm Q}_R(0)$ is the best characterised of the various phonon side-bands: it has a high peak value, and is the dominant source of absorption over a range of approximately $300\,$\percm\ around the peak. We have therefore used this band as a template for describing other phonon bands in the dataset. Specifically, the ${\rm Q}_R(0)$ band provides us with a description of the shape of a phonon band absorption; to match a particular phonon band we then scale the template according to the value of the peak absorption coefficient, and the location of that peak relative to the origin of the band (i.e. the zero-phonon line).

The phonon side-band shapes actually differ substantially amongst the various transitions in our data, and our procedure is just a simple heuristic which allows us to extrapolate from regions where the absorption coefficient is well-measured into regions where it is not.

In fact the ${\rm Q}_R(0)$ band itself required extrapolation in order to provide a template that is valid over a suitably large range of wavenumber. Our ${\rm Q}_R(0)$ data display approximately exponential decline in the absorption coefficient at large distances from the peak of the band, and we therefore used exponential functions to extend our phonon band template to wavenumbers smaller and larger than those for which we were able to measure the band shape.

\subsubsection{Data segments utilised}
The infrared absorptions of solid \ph\ are clustered around the various ${\rm S}_v(0)$ lines ($v=0,1,2\dots$), as is clear from figure 1 of Mengel, Winnewisser and Winnewisser (1998). Consequently we only utilised our data to define the imaginary part of the refractive index in those segments where the absorption signal is strong. Those segments are the following: 689-850$\,$\percm, 4{,}000-5{,}450$\,$\percm, 8{,}395-8{,}790$\,$\percm, and 8{,}970-8{,}974$\,$\percm. The data for each segment can be seen in figure 3. Outside of those segments (but within our instrumental window of 600-10{,}000$\,$\percm), our data were used indirectly. That is, the data were used to define models of the absorption spectrum, and we then employed the models in our refractive index description. The models are Lorentzian profiles, in the case of the ${\rm U_0(0)}$ and ${\rm S_2(0)}$ lines -- both of which are very narrow -- and phonon band absorptions as described in the previous section. 

\subsection{Long wavelength absorptions}
The pure rotational line, ${\rm S_0(0)}$, and the peak of the associated phonon branch, ${\rm S_R(0)}$, lie below the $600\;$\percm\ wavenumber limit of our spectrometer and are thus absent from our data. These transitions  are known to be strong absorptions of solid \ph, relative to the IR transitions we have measured, and it is therefore desirable to include a representation of these features in our description of the dielectric properties.

Our data show the ${\rm S_1(0)}$ line to be well represented by a Lorentzian absorption profile with FWHM of 0.084$\,$\percm, and we therefore represent the ${\rm S_0(0)}$ transition with a Lorentz Oscillator model (as detailed in \S3.1), yielding the same profile as the ${\rm S_1(0)}$ line. The wavenumber at resonance we use for the ${\rm S_0(0)}$ line is 355.7$\,$\percm\ (Buontempo et al 1982), and the integrated line-strength (Buontempo et al 1982; Okumura, Chan and Oka 1989), taken together with our profile shape, indicates a peak absorption coefficient of 159$\,$\percm.  

Unlike the zero-phonon ${\rm S_0(0)}$ line, the phonon branch, ${\rm S_R(0)}$, is not expected to follow a simple analytic form.  We have therefore chosen to represent the ${\rm S_R(0)}$ absorption profile by scaling the ${\rm Q_R(0)}$ phonon band, as described in \S2.5.4. Laboratory measurements  of the ${\rm S_R(0)}$ branch indicate that the peak absorption is 5.3$\,$\percm, and that the peak is approximately 55$\,$\percm\ above the ${\rm S_0(0)}$ line centroid (Buontempo et al 1982), and we scaled our template accordingly.

\subsection{Short wavelength absorptions}
There are, of course, rovibrational absorption bands of solid-\ph\ which lie at wavenumbers higher than we can measure with our spectrometer, corresponding to the second, third, fourth etc. vibrational overtones. As already noted, those transitions are relatively weak (Mengel, Winnewisser and Winnewisser 2013), and we make no attempt to include them in our dielectric model.

\section{Model dielectric function}
The strongest electromagnetic response of \htwo\ is to be found in the far-ultraviolet (FUV), where various absorption bands and the ionisation edge are all located. These FUV transitions are the dominant contributions to the dielectric function of solid \ph. Measurements of the FUV properties of solid \htwo\ were reported by Inoue, Kanzaki and Suga (1979), but there are several reasons why those results are difficult to use for our purpose here: they are only presented as optical densities in graphical form; the sample thickness was not determined; and the {\it ortho/para\/} content of the samples is unspecified. Consequently we are obliged to construct a model of the FUV absorptions based on individual molecules of \htwo, and scaled to the density of the solid. We first treat the contributions of the discrete transitions, and then deal separately with the bound-free absorptions.

\subsection{FUV absorption lines}
The lowest-lying absorption bands are the Lyman and Werner systems; these are transitions from the electronic ground-state to the excited electronic states $B(^{1\,}\!\Sigma_u^+)$ and $C(^{1\,}\!\Pi_u)$ respectively (Field, Somerville and Dressler 1966).  At the low temperatures of interest to us, essentially all of the \htwo\ molecules are expected to be in the ground rovibrational level of the ground electronic state,  and the only relevant transitions are those which connect to this state. This limits us to the ${\rm R(0)}$-branch transitions of each absorption band\footnote{In contrast to the IR transitions of \htwo, which are quadrupolar, the dominant UV transitions are electric dipole, so $\Delta J=1$.} -- i.e. the rotational quantum number of the upper state must be $J=1$ -- but there is no restriction on the vibrational quantum number for the upper level. Each of these lines can be represented by a Lorentz oscillator, so their combined contribution to the susceptibility for radiation of angular frequency $\omega$ can be modelled as
\begin{equation}
\chi_{{}_{lo}}(\omega)={{ne^2}\over{\epsilon_0m_e}}\sum_j{{f_j}\over{\omega_j^2-\omega^2-i\,\omega\gamma_j}},
\end{equation}
(e.g. Fox 2012) where the subscript $j$ identifies the contributing transition, for which $f_j$ is the oscillator strength, $\omega_j$ the angular frequency at resonance, and $\gamma_j$ the damping constant. As usual,  $e$, $m_e$ and $\epsilon_0$ denote the charge and mass of the electron, and the permittivity of free-space. The quantity $n$ is the number of molecules per unit volume. 

To make use of equation (2) we need values of $\{\omega_j, f_j, \gamma_j\}$. We have compiled suitable data from the literature, and in Appendix A we list values of these quantities for each of the transitions included in our model. In summary we have 99 transitions, made up of: 38 transitions in the Lyman Band; 14 transitions in the Werner Band; and 47 transitions involving a further 5 excited electronic states.

The combined oscillator strength of all transitions should equal the number of electrons in each molecule, i.e. 2, whereas the discrete transitions in our model contribute $\sum_jf_j\simeq0.829$. Certainly our list of discrete transitions is incomplete, but the major part of the discrepancy is due to bound-free transitions. Because they form a spectral continuum, these transitions are not naturally described in terms of Lorentz oscillators.

\subsection{Bound-Free continuum}
To model the bound-free continuum we use a piecewise analytic fit (Yan, Sadeghpour and Dalgarno 1998), to the measured bound-free cross-section of the \htwo\ molecule, $\sigma_{bf}(\omega)$ (Samson and Haddad 1994). Near the ionisation edge the fit we employ is the modified one due to Liu and Shemansky (2004). For low number-densities of molecules, the imaginary part of the bound-free contribution to the refractive index, ${\rm n}_{bf}(\omega)$, is then given by ${\rm Im\{n}_{bf}(\omega)\}=n\, \sigma_{bf}(\omega)c/2\omega$. The corresponding real part can be obtained from the imaginary part by applying the appropriate Kramers-Kronig relation --- as per Draine and Lee (1984), for example. Knowing both real and imaginary parts of the refractive index contribution, we can determine the bound-free susceptibility contribution from $\chi_{{}_{bf}}={\rm n}_{bf}^2-1$.

The total electromagnetic response of a dilute collection of \htwo\ molecules, due to both line and continuum interactions, is then just
\begin{equation}
\chi_{{}_d}(\omega)=\chi_{{}_{lo}}(\omega)+\chi_{{}_{bf}}(\omega).
\end{equation}
We have not yet specified the number density of molecules, $n$; that is fixed by the mass density of the solid ($0.087\,{\rm g\,cm^{-3}}$). In gas-phase the dielectric constant scales linearly with $n$, so that  $\epsilon=1+\chi_{{}_d}$. But in the solid phase the packing is so close that the polarisation of neighbouring molecules has a significant effect on the electric field at any given lattice site, and $\chi=\epsilon-1\ne\chi_{{}_d}$. We model this in the usual way, utilising the Clausius-Mossotti/Lorenz-Lorentz relation, which yields
\begin{equation}
\epsilon(\omega)=1 + {{3\chi_{{}_d}(\omega)}\over{3-\chi_{{}_d}(\omega)}}
\end{equation}
for the solid. Strictly the Clausius-Mossotti/Lorenz-Lorentz relation is only valid for cubic crystals, but the preferred hcp crystal of solid \ph\ is, as noted earlier, very similar to a fcc structure, so we do not expect to incur a large error in utilising the relationship here. \alert{(A rough estimate of the magnitude of the error can be obtained by analogy with solid ${}^4$He, which also forms an hcp crystal at low temperatures. For that crystal the optical birefringence (Vos et al 1967; Kronig and Sonnen 1958), indicates a departure from equation (4) that is $\sim10^{-4}$ of $\epsilon-1$. This is much smaller than the systematic errors involved in our estimate of $\chi_{{}_d}$.)}

\subsection{Infrared absorptions}
Our model of the dielectric function expressed by equations 3 and 4 does not include the contribution due to infrared transitions. As we have measured the IR absorptions (\S2), we can determine the contribution of the IR transitions to $\epsilon$ in a manner closely analogous to our determination of $\chi_{{}_{bf}}$ in \S3.2. In particular we have 
\begin{equation}
{\rm Im\{n}_{ir}\}={{\alpha\lambda}\over{4\pi}}
\end{equation}
\alert{(where $\lambda$ is the vacuum wavelength),} then ${\rm Re\{n}_{ir}\}$ is determined from ${\rm Im\{n}_{ir}\}$ by applying the relevant Kramers-Kronig transform, and the susceptibility contribution is $\chi_{{}_{ir}}={\rm n}_{ir}^2-1$.

Because we determined the IR absorption coefficient of the solid itself, rather than a dilute collection of molecules, it is not necessary to apply a local-field correction to the IR susceptibility. Our model dielectric constant therefore follows from adding $\chi_{{}_{ir}}$ to equation 4:
\begin{equation}
\epsilon(\omega)=1 + {{3\chi_{{}_d}(\omega)}\over{3-\chi_{{}_d}(\omega)}} + \chi_{{}_{ir}}(\omega),
\end{equation}
where $\chi_{{}_d}$ is again given by equation 3. The resulting refractive index is shown in figure 4. The same information is given in table 1 as $\epsilon-1$ versus $\lambda^{-1}$; this table is very large and is provided in electronic form.

\begin{table}
\caption{The model dielectric constant of solid \ph\  as a function of wavenumber, running over wavelengths from one metre down to slightly below one \AA ngstrom. Only the first-two and last-two lines are shown here; the full table is available in electronic form.}
\begin{tabular}{@{}ccc}
\hline
$1/\lambda$ &  ${\rm Re\{\epsilon-1\}}$  & ${\rm Im\{\epsilon-1\}}$       \\
$({\rm cm^{-1}})$   &        &   \\
\hline
$1.00000\times10^{-2}$ 		&   $0.269283$	& $1.00205\times10^{-13}$\\
$1.10975\times10^{-2}$ 		&   $0.269283$	& $1.11202\times10^{-13}$\\
$\dots$  & $\dots$ & $\dots$\\  
\\
$\dots$  & $\dots$ & $\dots$\\  
$9.93305\times10^{7}$ 		&   $-4.49018\times10^{-7}$	& $2.69114\times10^{-13}$\\
$1.01644\times10^{8}$ 		&   $-4.28809\times10^{-7}$	& $2.42858\times10^{-13}$\\
\end{tabular}
\medskip
\end{table}

\begin{figure}
\includegraphics[width=85mm]{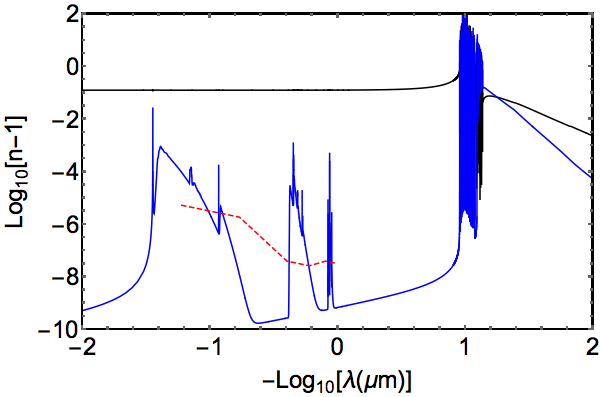}
\vskip-0truecm
\includegraphics[width=85mm]{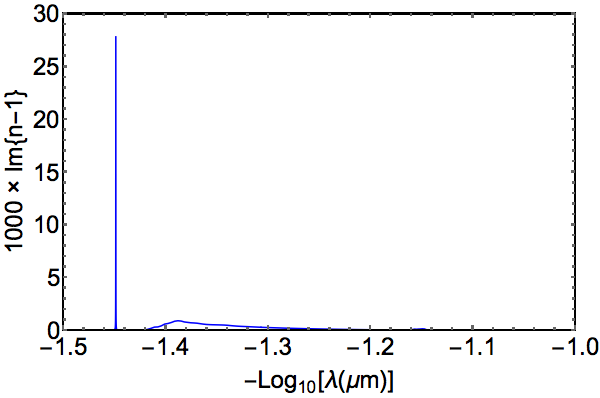}
\vskip-0truecm
\includegraphics[width=85mm]{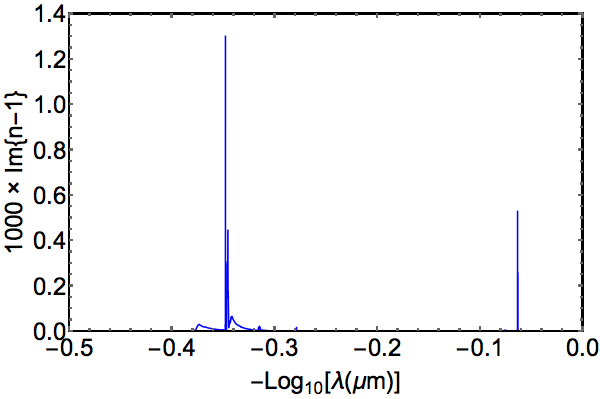}
\vskip-0.1truecm
\caption{\alert{Real (black) and imaginary (blue) parts of the refractive index (minus 1) of solid \ph, as a function of wavenumber (top panel).} Our laboratory data lie within the region $1.2\la-\log_{10}\lambda({\rm\mu m})\le0$, and absorptions lying outside this region have been modelled with the aid of published data. At low wavenumbers the real part of $\sqrt{\epsilon}-1$ is positive, whereas at high wavenumbers it is negative, and in the absorption bands around $10\;{\rm\mu m}^{-1}$ it changes sign many times; we have therefore plotted the absolute value. Between 11 and 14.5~eV ($0.95\la-\log_{10}\lambda({\rm\mu m})\la1.07$) discrete transitions, detailed in Appendix A, are the main contribution to absorption. The bound-free continuum dominates above the $\simeq$15.4~eV ($-\log_{10}\lambda({\rm\mu m})\simeq1.09$) ionisation edge. The red dashed line shows a rough confidence level for ${\rm Im\{n-1\}}$: below this line the absorption coefficient is not well characterised. \alert{The middle and bottom panels show the pure rotational and the fundamental vibrational band absorptions, respectively, on a linear scale.}}
\end{figure}

\subsection{Comparison of model with data}
Some quantitative checks on the accuracy of our model are available. First, the (real part of the) dielectric constant of solid \ph\ at low frequency (kHz regime) has been measured to be $\epsilon(0)-1=0.2833$ (Constable, Clark and Gaines 1975), whereas our model predicts 0.2693. Secondly, measurements of the (real part of the) refractive index of solid \ph\ in the visible and near-infrared bands yield values in the range $0.13\la\sqrt{\epsilon}-1\la0.14$ (Perera et al 2011), whereas our model yields 0.127-0.132 across the same region of the spectrum. Thirdly, in the hard X-ray region the binding energy of the electrons is negligible and the refractive index should be just that of a cold plasma of electron density $2n$, whereas our model tends towards a cold plasma with density $1.894n$. This is another way of saying that we have captured almost all of the oscillator strength. On all three of these tests our model performs well; it systematically under-predicts the response of the medium at all frequencies, but only by about 5\%.

In comparison with the data of Inoue, Kanzaki and Suga (1979), our model for the UV response of solid \ph\ displays discrete transitions which are much too sharp. That is because our model for the UV is based on a dilute collection of \htwo\ molecules, but scaled to the density of the solid. As such the model ignores the intermolecular interactions, occuring in the solid, that act to broaden the transitions. \alert{We reiterate, therefore, that our model of the UV resonances should not be relied on within the UV band itself --- it is a good approximation only at much lower and much higher frequencies.}

\section{Discussion}
Compared to the dielectric constants of conventional interstellar dust materials -- i.e. ``astronomical silicate'' and graphite (e.g. Draine and Lee 1984) -- the most striking aspect of figure 4 is the weakness of the absorption at energies below the Lyman band. Because most stars put out most of their power in that region, it is immediately clear that \alert{any pure hydrogen dust in the diffuse interstellar medium will be subject to comparatively little radiative heating.} At present it is unclear how much the absorption evident in figure 4 contributes to heating of the crystal, as the elementary molecular rovibrational excitations are long-lived and de-excitation could be principally radiative. Certainly the phonon branches contribute heat immediately on absorption of a photon, but the phonons themselves carry off an average $\la10^2\,$\percm\ per absorption, and that is small compared to the energy of the absorbed photon for most of the spectral range in figure 4.

At low wavenumbers the real part of the refractive index of solid \htwo, although large compared to the imaginary part, is also much smaller than that of conventional grain materials (e.g. Draine and Lee 1984). This means that the wavelength of peak extinction for a pure hydrogen grain is expected to be shorter than that for a silicate or graphite grain of the same size.

\alert{In addition to the possibility of pure solid \htwo\ grains, our results will be useful for calculating the optical response of composite grains consisting, for example, of thick \htwo\ mantles on cores of other materials -- refractory or icy. Although the \htwo\ mantles themselves would not absorb much starlight, the core material may absorb quite strongly, thus heating the whole grain. The radiative heating of such grains in the diffuse ISM may thus be too high for an \htwo\ mantle to persist, and consequently grains of this type are likely to be of interest primarily in the context of dense interstellar clouds.}

One of our motivations for precise characterisation of the infrared dielectric constant of solid \ph\ was the possibility of identifying a spectral signature by which this material might be recognised in interstellar dust. In principle we have achieved that, as the spectral shape of the absorption coefficient in figure 3 is really quite distinctive. However, these absorptions are very weak and the imaginary part of the refractive index is small compared to the real part, at all wavenumbers below the Lyman band. That means that it will be difficult to use these IR features to test whether interstellar dust contains solid \htwo, because the extinction curve should be dominated by scattering and is thus expected to be relatively featureless (upper panel in figure 4). For example: the strongest line in our data is ${\rm S_1(0)}$, and in the vicinity of that line the real part of ${\rm n}-1$ varies by less than 2\% of its local value. Furthermore that fluctuation is confined to within $\pm0.1\,$\percm\ around the line, so even the strongest IR absorptions would be difficult to detect in the extinction curve. 

\section{Conclusions}
We have characterised the dielectric constant of solid $para$-\htwo. \alert{Our description is based on laboratory measurements of the absorption coefficient of the solid in the wavelength range $1<\lambda({\rm\mu m})<16.6$, together with published data on the low-frequency absorption spectrum of the solid and published data on the UV absorptions of the \htwo\ molecule.} The real part of the dielectric constant is determined from the imaginary part by using the Kramers-Kronig relations. Because the UV transitions had to be modelled on the basis of absorptions measured in gas phase, the spectral structure in the vicinity of the  UV resonances is not to be trusted. At higher photon energies, where the dielectric properties resemble a free-electron plasma, the optical constants appear to be good to $\sim5$\%. Comparable accuracy is demonstrated for the real part of the dielectric constant at frequencies well below the UV resonances. The accuracy of our own measurements of the absorption coefficient is also limited, by sample thickness determination, to $\sim5$\% in regions of strong absorption, but is much poorer in regions which are highly transparent.

The infrared absorption spectrum of solid \ph\ is very distinctive, with a great variety of line shapes and widths, and exhibits transitions which are unique to hydrogen in the condensed phase. However, all infrared absorptions are weak and consequently the imaginary part of the refractive index is small compared to the real part. \alert{We therefore expect that solid \ph\ contributes to extinction primarily via a featureless scattering continuum, making it difficult to recognise hydrogen dust via the IR absorptions of the pure matrix.}

\section*{Acknowledgments}
DTA and SCK thank the Wyoming NASA EPSCoR Material Science and Engineering program for a Faculty Research Grant and a Graduate Research Fellowship, respectively. MAW thanks Oxford Astrophysics for hospitality.

\appendix
\section{Discrete transition list}
In this appendix we provide information on the data used in our Lorentz oscillators model of the discrete transitions of \htwo, described in \S3 of this paper. We supply the data themselves as a list of transitions, in the form of an electronic supplement, with the following properties.

There are 99 lines in the list, each one is from the ${\rm R(0)}$ branch, and for each line there are five pieces of data. The supplement is thus a table with 5 columns and 99 rows.  The columns are: (1) the index number we assigned to identify the line; (2) the vibrational quantum number, $v$, of the upper level; (3) the wavenumber of the transition, in ${\rm cm^{-1}}$; (4) the Einstein $A_{21}$ coefficient of the transition, in MHz; and (5) the total decay rate, $\Gamma$, for the upper level, in MHz. The ordering of the lines is in groups according to the excited electronic state of the transition, and within each group the ordering is by increasing values of $v$. The groups are as follows:

Lines 1-38: state $B(^{1\,}\!\Sigma_u^+)$, known as the Lyman Band. Wavenumbers and $A_{21}$ coefficients taken from Abgrall et al (1993a), except for  $v = 23, 33$ and $v = 35-37$ which were taken from Abgrall and Roueff (1989). All values of $\Gamma$ were taken from Abgrall, Roueff and Drira (2000).

Lines 39-52: state $C(^{1\,}\!\Pi_u)$, known as the Werner Band. Wavenumbers and $A_{21}$ coefficients taken from Abgrall et al (1993b), except for  $v = 13$ which was taken from Abgrall and Roueff (1989). All values of $\Gamma$ were taken from Abgrall, Roueff and Drira (2000).

Lines 53-59: state $B^\prime(^{1\,}\!\Sigma_u^+)$. Wavenumbers and $A_{21}$ coefficients taken from Abgrall et al (1994). All values of $\Gamma$ were taken from  Abgrall, Roueff and Drira (2000).

Lines 60-62: state $D(^{1\,}\!\Pi_u)$. Wavenumbers and $A_{21}$ coefficients taken from Abgrall et al (1994). All values of $\Gamma$ were taken from  Abgrall, Roueff and Drira (2000).

Lines 63-77: state $B^{\prime\prime}\bar{B}(^{1\,}\!\Sigma_u^+)$. All data taken from Glass-Maujean et al (2007). Note that this electronic state exhibits a double-well structure; the vibrational quantum number given in our line list corresponds to the outer well.

Lines 78-91: state $D^\prime(^{1\,}\!\Pi_u)$. All data are from Glass-Maujean et al (2008a). Note that the total decay rate for the levels $v=3-9$ was smaller than could be measured by Glass-Maujean et al (2008a). We have assumed the value $10^4$~MHz for each of these levels, which is a factor of 12 smaller than the lowest measured value for this state.

Lines 92-99: state $5p\sigma(^1\Sigma_u^+)$. All data are from Glass-Maujean et al (2008b). Note that we adopt their calculated value of $A_{21}$ for $v=3$, as the observed line is blended, and for the width of this line we have used the value they list for $P(2)$, $v=3$, which corresponds to the same upper level.

Table A.1 reproduces the first-two and last-two lines in the list. The various quantities listed in the table are related to the quantities employed in the Lorentz oscillator model thus:
\begin{equation}
\omega_j=2\pi{c\over\lambda},
\end{equation}
\begin{equation}
f_j =  {3\over{4\pi}} A_{21}  {{\lambda^2 m_e} \over{\alpha_{f} h}} ,
\end{equation}
where $\alpha_{f}\simeq1/137$ is the Fine Structure Constant, and $h$ is Planck's Constant. The damping constant is simply
\begin{equation}
\gamma_j={1\over2}\Gamma.
\end{equation}

\begin{table}
\caption{The first-two and last-two lines in the electronic table of transitions used in the Lorentz oscillator model of \S3. \alert{The complete table is available in electronic form.}}
\begin{tabular}{@{}lcccc}
\hline
Line  & $v$ & $1/\lambda$ &  $A_{21}$                          & $\Gamma$                   \\
\#     &        &$({\rm cm^{-1}})$   &    (MHz)         & (MHz) \\
\hline
1 &   0       &   90242.36    &   3              &   1,900   \\
2 &   1       &   91558.74    &   11  &   1,700    \\
$\dots$  & $\dots$  & $\dots$  & $\dots$ & $\dots$\\  
\\
$\dots$  & $\dots$  & $\dots$  & $\dots$ & $\dots$\\  
98 &   8     &   133141       &   3.1  &   15,000 \\
99 &   9     &   134023       &   5.1  &   330,000 \\
\end{tabular}
\medskip
\end{table}

\label{lastpage}
\end{document}